\documentclass[runningheads]{llncs}
\usepackage[T1]{fontenc}
\usepackage{graphicx}
\usepackage[table,xcdraw]{xcolor}
\usepackage{booktabs}
\usepackage{orcidlink}

\begin{document}
\title{Large-area Spatially Aligned Anchors}

\author{Joni Vanherck\inst{1}\orcidlink{0009-0008-1312-0382} \and Brent Zoomers\inst{1}\orcidlink{0009-0000-7071-8534} \and
Lode Jorissen\inst{1}\orcidlink{0000-0001-7228-9315} \and  Isjtar Vandebroeck\inst{2}\orcidlink{0000-0002-1619-7368} \and Eric Joris\inst{2}\orcidlink{0000-0002-1533-3014} \and
Nick Michiels\inst{1}\orcidlink{0000-0002-7047-5867}}
\authorrunning{J. Vanherck et al.}
\institute{Hasselt University - Flanders Make - Expertise Centre for Digital Media, Diepenbeek, Belgium \\
\email{\{firstname.lastname\}@uhasselt.be} \and
CREW - CREW Brussels, Vandernootstraat 23/ 8, Sint-Jans-Molenbeek, Belgium \\
\email{isjtar@crew.brussels, eric.joris@crew.brussels}}
\maketitle            
\begin{abstract}
Extended Reality (XR) technologies, including Virtual Reality (VR) and Augmented Reality (AR), offer immersive experiences merging digital content with the real world. Achieving precise spatial tracking over large areas is a critical challenge in XR development. This paper addresses the drift issue, caused by small errors accumulating over time leading to a discrepancy between the real and virtual worlds. Tackling this issue is crucial for co-located XR experiences where virtual and physical elements interact seamlessly. Building upon the locally accurate spatial anchors, we propose a solution that extends this accuracy to larger areas by exploiting an external, drift-corrected tracking method as a ground truth. During the preparation stage, anchors are placed inside the headset and inside the external tracking method simultaneously, yielding 3D-3D correspondences. Both anchor clouds, and thus tracking methods, are aligned using a suitable cloud registration method during the operational stage. Our method enhances user comfort and mobility by leveraging the headset's built-in tracking capabilities during the operational stage, allowing standalone functionality. Additionally, this method can be used with any XR headset that supports spatial anchors and with any drift-free external tracking method. Empirical evaluation demonstrates the system's effectiveness in aligning virtual content with the real world and expanding the accurate tracking area. In addition, the alignment is evaluated by comparing the camera poses of both tracking methods. This approach may benefit a wide range of industries and applications, including manufacturing and construction, education, and entertainment. 

\end{abstract}

\section{Introduction}
\begin{figure}
    \centering
    \includegraphics[width=0.8\textwidth]{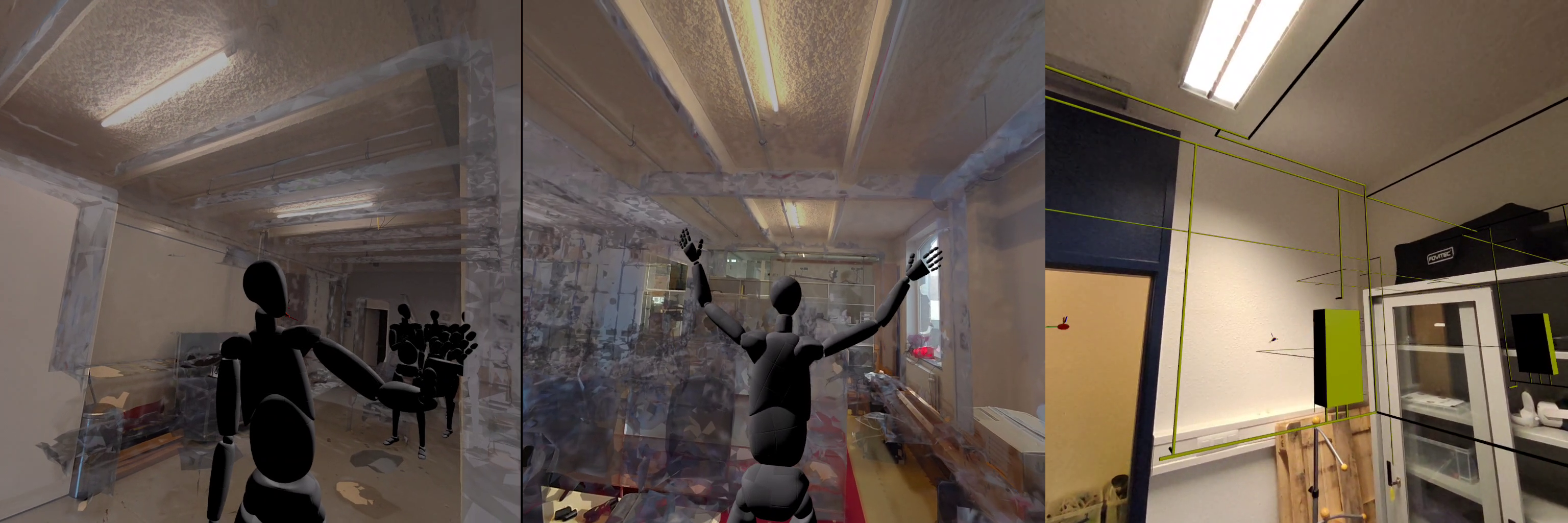}
    \caption{Two example applications leveraging content alignment with the real world. The left and middle images show an artistic performance, while the right image shows an application that may help with utility infrastructure maintenance. These images represent the viewpoints of the user, using the passthrough mode of the Quest 3.  }
    \label{fig_crew_img}
\end{figure}

Extended Reality (XR) technologies, encompassing Virtual Reality (VR) and Augmented Reality (AR), are becoming increasingly popular for creating immersive experiences that blend the real world with digital content. From entertainment and gaming to education, healthcare, and manufacturing, VR and AR offer unique opportunities to immerse users in digital environments and augment their perception of the real world. However, as these technologies continue to evolve, the challenge of achieving precise and consistent spatial tracking over large areas remains a critical challenge. Accomplishing this in a practical manner adds further complexity. 
Using our suggested solution, Figure \ref{fig_crew_img} presents two practical examples demonstrating how virtual scenes seamlessly integrate with reality. One showcases an artistic performance, while the other features utility infrastructure maintenance.

Two tracking paradigms exist: outside-in tracking and inside-out tracking. With outside-in tracking, external devices/sensors such as cameras and optical base stations determine the pose of items of interest. Although outside-in tracking is very accurate, these technologies are often expensive and difficult to scale to large areas. On the other hand, inside-out tracking, which utilizes the cameras on the device itself, presents a more scalable solution that enhances user mobility without the need for additional sensors. However, the accuracy of these systems is limited by drift~\cite{mcgill_quest_2020}, caused by small errors accumulating over time. For larger areas, the drift will become too large to achieve co-location. Other techniques try to solve this by adding many markers and hardware sensors~\cite{podkosova_immersivedeck_2016}, which are both impractical. Preferably, this is solved within the standalone headset itself, by adding only a few to no markers at all. 

The objective of this paper is to address the issue of drift within the headset. Spatial anchors~\cite{thetuvix_spatial_2023} mitigate this problem on a small scale by correcting drift on self-defined reference points within a limited area. The spatial anchors are most effective within an area of 3 m from the origin of the anchors. The aim of this paper is to extend this accuracy to larger areas. The proposed method achieves this by leveraging spatial anchors and an external, drift-corrected tracking method to establish anchors both externally and within the headset. These anchors serve as reference points to align the tracking methods with each other. 

Accurate tracking is crucial to integrate virtual and real-world elements in co-located XR experiences seamlessly. Virtual and physical elements coexist in co-located environments, allowing users to interact with both simultaneously. Ensuring a consistent mapping between the physical and virtual positions is necessary to prevent user collisions with the physical environment and accurately align AR content. Another benefit involves content authoring. Creating content becomes more challenging when using spatial anchors because each anchor has its own coordinate system. Also, spatial anchors are continuously being corrected for drift, which means the relation between spatial anchors changes, further complicating the matter. In contrast, our approach simplifies content authoring by providing one global system for building the scene.  

This paper was inspired by the system introduced by Michiels et al.~\cite{michiels_tracking_2024}, which also addresses the drift issue. However, a drawback of their approach is that the external tracking camera remains mounted to the headset the whole time, increasing the headset's weight and thus, causing additional discomfort for users. Additionally, a backpack is required since the tracking method is executed on an external PC. Because tracking operates at a lower framerate, more complex sensor fusion is required. Our proposed method overcomes these problems by utilizing the headset's tracking capabilities during the operational stage, enabling standalone functionality and enhancing user freedom. In addition, our proposed method introduces a decoupling of the external tracking method and the headset's tracking capabilities, allowing for the integration of any accurate external tracking method with existing XR headsets that support spatial anchors.

The method is empirically evaluated by overlaying a mesh of the environment on the passthrough layer of the headset. The accuracy of the alignment can be easily noticed. Additionally, the system is evaluated against the underlying external tracking method by comparing the corresponding camera poses. Our system is as practical as using default spatial anchors, but increases the range of the accurate tracking area from 3 meters. The code is publicly available on GitHub: \url{https://github.com/EDM-Research/UE-LASAA}. 

\section{Related work}
Numerous systems are designed with precise outside-in tracking capabilities. Outside-in tracking employs external measurement devices to track a user inside a room. These devices may be cameras~\cite{furtado_comparative_2019}, projective beacons~\cite{niehorster_accuracy_2017} or even acoustic trackers~\cite{mao_cat_2016}. Despite their accuracy, these systems are expensive and annoying to configure when scaling to large environments. 

Another approach for user tracking is inside-out tracking. With inside-out tracking, on-device sensors, such as cameras and IMU sensors, are used for tracking. The advancement of inside-out tracking techniques has made seamlessly integrating room-scale consumer devices like the Quest headsets, the HoloLens, and the Magic Leap affordable. Several IMU-based tracking methods have been researched, both completely relying on IMU sensor data \cite{bai_high-precision_2020} and combined with other sensors such as cameras using sensor fusion \cite{alatise_pose_2017}. The most common visual technology within this category is SLAM~\cite{durrant-whyte_simultaneous_2006}, which tries to simultaneously find a user's location while also constructing a map of the environment. Due to sensor noise and the limited precision of computing devices, small errors accumulate over time. This issue is called drift, limiting the accuracy of tracking. For example, whenever users revisit a place they already saw, there might be a gap between the real and virtual locations of users. Drift impacts most inside-out tracking methods, including IMU-based techniques. Solutions like loop-closing~\cite{mur-artal_fast_2014} have been proposed to reduce the impact of drift. Recent advancements employ Gaussian splatting for SLAM: SplaTAM~\cite{keetha_splatam_2024} and Gaussian SLAM~\cite{yugay_gaussian-slam_2024}. Also, visual-inertial SLAM solutions exist \cite{campos_orb-slam3_2021,song_dynavins_2022} that combine the visual and inertial information for SLAM tracking. These SLAM approaches can be integrated into cheap and untethered portable devices. However, drift prevents these solutions from scaling properly to larger environments. McGill et al.~\cite{mcgill_quest_2020} showed that the Oculus Quest is only accurate to $\pm0.4$ m in a small environment of 6 m², with increasing drift over time, especially when scaling to large areas. Although new-generation XR headsets improve tracking accuracy, they are not flawless. 

SLAM methods rely on natural image features for tracking. When possible, artificial image features consisting of ArUco codes~\cite{garrido-jurado_automatic_2014} or other custom markers can be introduced, similar to the approach taken by Podkosova et al.~\cite{podkosova_immersivedeck_2016}. These markers allow more accurate tracking since they can be used as real-world static reference points, thereby removing drift entirely. However, setting up an environment with a large number of markers is not always desired, mainly because it is impractical. Also, the markers should be visible at all times for tracking purposes. Small errors in pose estimation can still occur due to different factors such as inaccurate detection, markers not being completely flat, etc. 

Apart from loop-closing, other methods that reduce drift have been researched: Spatial Anchors~\cite{thetuvix_spatial_2023} and World Locking tools (WLT)~\cite{vtieto_world_nodate}. Spatial anchors are reference points in the physical world that are tracked and corrected for drift. Objects attached to a spatial anchor are continuously moving to adjust their pose to account for drift. Each spatial anchor does this independently of other anchors, which means they do not have a rigid relationship. This complicates content creation since the content must be placed per spatial anchor, and there is no fixed relation between the anchors. This approach to drift correction is only accurate within 3 m of the anchor~\cite{thetuvix_spatial_2023}. Spatial anchors can be shared across multiple devices, easily achieving multi-user experiences.
Microsoft's World Locking Tools tries to scale this method to larger areas utilizing spatial anchors and space pins. Space pins are reference points in the physical world with a known pose, allowing us to align the real world's coordinate system with the virtual world's coordinate system. These points can be QR codes encoding additional information. This setup requires manually adjusting the virtual poses of the game engine's space pins to match the markers' physical layout. WLT adjusts the headset's pose in relation to several anchors to correct drift for the whole environment. As shown by Teruggi and Fassi~\cite{teruggi_mixed_2022}, WLT is accurate in the order of 15 cm in large environments of 40 $\times$ 85 m. 

We try to reduce drift inside the headset by using the headset's tracking capabilities and spatial anchors combined with any external, drift-corrected tracking method. This approach reduces drift inside the headset while allowing standalone functionality. The proposed method works with any headset that supports spatial anchors. While our approach is similar to WLT, it is more flexible and easier to set up, and works with any headset that supports spatial anchors. Depending on the external tracking method that is used, artificial features are not necessary, contrary to WLT. Finally, this framework will be publicly available and can be easily used as a plugin in Unreal Engine 5.

\section{Methodology overview}
\begin{figure}[h]
    \centering
    \includegraphics[width=\textwidth]{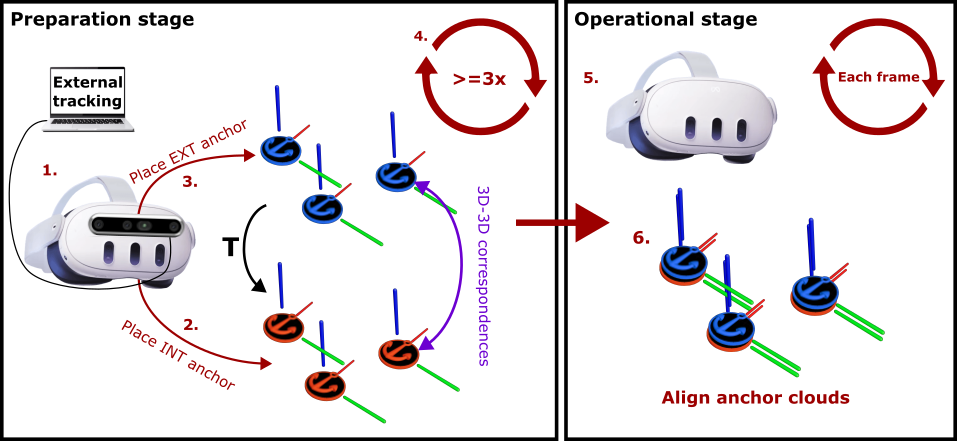}
    \caption{A schematic overview of the proposed method. \textbf{1.} The external tracking method is configured. \textbf{2.} Place an INT anchor. \textbf{3.} Simultaneously, save an EXT anchor. \textbf{4.} Repeat steps 2-3 at least 3 times, until a sufficient area of the scene is covered. \textbf{5.} Unmount/disable the external tracking method.  \textbf{6.} Align both anchor clouds in every frame using a suitable cloud registration method. This is done by finding and applying transformation T.}
    \label{fig_overview}
\end{figure}

The methodology consists of two stages: a preparation stage and an operational stage. In the preparation stage, the user walks around with an external tracking camera mounted on the headset, placing anchors (3D reference points) in the tracking environment. During this stage, two tracking methods are used simultaneously: the head-mounted display tracking and an external tracking method. The additional camera and external tracking method can be discarded during the operational stage, and the system works in standalone mode. A schematic overview is shown in Figure \ref{fig_overview}.
There are two tracking methods: the head-mounted display's (internal) tracking and the external, drift-corrected tracking method. For each technique, anchors will be saved. The spatial anchors inside the HMD will be called INT (internal) anchors, while the anchors in the external method will be called EXT (external) anchors. These anchors are in the same coordinate system as the content model. 

The proposed method works as follows: 
\begin{enumerate}
    \item Set up the external tracking method and mount the camera to the headset.
    \item Place an INT anchor in the headset coordinate system where many features are inside the room. 
    \item Simultaneously, save an EXT anchor of the same pose in the external coordinate system. This creates 3D correspondences between both tracking methods. 
    \item Repeat steps 2-3 at least three times until a sufficient area of the scene is covered with anchors.
    \item Unmount the extra tracking camera and use the headset standalone. 
    \item Align the EXT anchors with the INT anchors in every frame using a suitable cloud registration method. This is achieved by finding and applying transformation T, which is the transformation from the external to the internal tracking space. 
\end{enumerate}

\section{Preparation stage}
During the preparation stage, all information is gathered to align the virtual content with the real world accurately. Correspondences between the tracking methods are obtained by simultaneously employing the internal and external tracking methods and placing anchor points in both tracking environments. 
\subsection{External camera calibration}
The INT anchors in the headset are placed at the HMD's position, while the EXT anchors are set at the external camera's position. We need to calibrate the camera to the headset to adjust for this difference and find the relative transformation. This is done using two methods: the controller-based method described by Michiels et al.~\cite{michiels_tracking_2024} and a method solely based on the placement of the anchors. The accuracy of this extrinsic calibration is essential as it impacts the accuracy of the complete system.  
\subsubsection{Controller-based calibration method}
\begin{figure}[h]
    \centering
    \includegraphics[width=\textwidth]{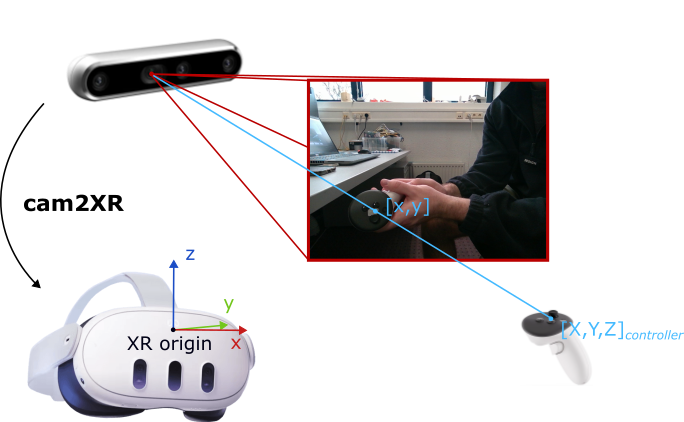}
    \caption{The calibration of the external camera to the XR headset, exploiting the position of the controller.}
    \label{fig_xr_calib}
\end{figure}

 The camera calibration relies on the known position of the controller in relation to the headset. By capturing the known 3D positions of the controllers and their corresponding 2D projections in the camera image, the pose of the external camera in relation to the XR headset ($\mathrm{cam2XR}$) can be determined by solving the following equation:
\begin{equation}
[x,y] = P_\mathrm{cam} \times \mathrm{cam2XR}^{-1} \times [X,Y,Z]_\mathrm{controller}
\end{equation} where $P_\mathrm{cam}$ is the projection matrix of the external camera, $[x,y]$ is the pixel coordinate of the origin of the controller, and $[X,Y,Z]_\mathrm{controller}$ is the 3D position of the controller. This calibration method performs well, but is cumbersome to execute. The process is shown in Figure \ref{fig_xr_calib}.

\subsubsection{Anchor-based calibration method}
\begin{figure}[h]
    \centering
    \includegraphics[width=\textwidth]{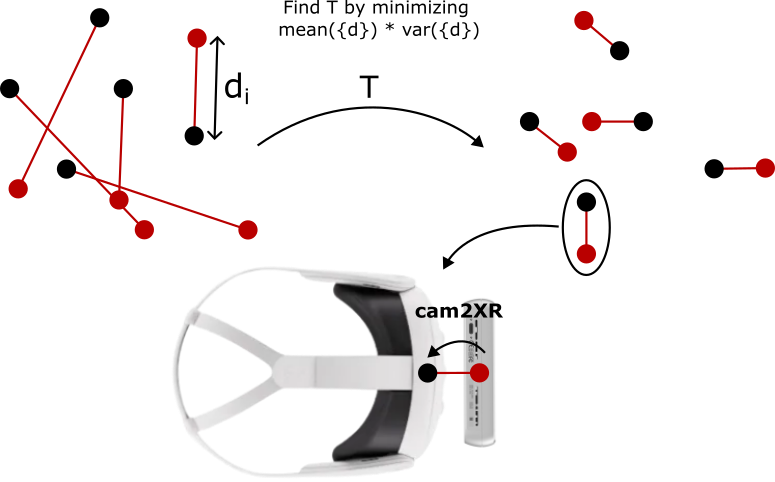}
    \caption{The calibration method of the external camera to the XR headset, exploiting the poses of the anchors. There are two anchor clouds with 3D-3D correspondences. These clouds are not aligned due to an unknown transformation T: the external tracking origin to the internal tracking origin. This transformation can be found by minimizing the cost function: $\mathrm{cost} = \mathrm{mean}(\{d\}) * \mathrm{var}(\{d\})$, where $\{d\}$ is the distance between anchor correspondences. Once this transformation has been determined, the extrinsic parameters $\mathrm{cam2XR}$ can be found.}
    \label{fig_anchorCalibrate}
\end{figure}

Figure \ref{fig_anchorCalibrate} illustrates the calibration method leveraging the anchor poses. To justify this approach, we assume there is minimal drift within a small space using spatial anchors. As already mentioned, spatial anchors are accurate within 3 meters. While the calibration of the external camera to the headset remains a distinct preprocessing step, this method offers simplicity thanks to the ease of placing anchors, which is a familiar process in subsequent stages. Although any external tracking method can be utilized, employing a drift-free method like marker-based tracking improves the accuracy.

Initially, several anchors are placed in a small range. It is crucial that these anchors are placed in various directions to provide enough constraints for the convergence of the algorithm. Two anchor clouds, internal and external, are established with known 3D-3D correspondences that are distinct due to the camera-to-headset relation. However, the clouds are not aligned correctly due to an unknown transformation T, which coincides with the transformation from external to internal space. This transformation will also be adjusted for continuously during the operational stage. This transformation is determined by minimizing a cost function:
\begin{equation}
    \mathrm{cost} = \mathrm{mean}(\{d\}) * \mathrm{var}(\{d\})
\end{equation} where $\{d\}$ is the distance between the 3D-3D correspondences. 
The distance between anchor correspondences should remain constant since the camera is fixed to the headset. This consistency is achieved by minimizing the variance of the distance between anchors.
This cost function is minimized using the Levenberg-Marquardt non-linear optimization technique~\cite{more_levenberg-marquardt_1978}. Subsequently, after applying the determined transformation, the calibration of the camera to the headset ($\mathrm{cam2XR}$) can be easily derived from the anchor pairs. The average transformation is determined to reduce the impact of outliers. This process is shown in \ref{fig_anchorCalibrate}.

\subsection{External tracking method}
In this paper two external tracking methods are demonstrated: marker-based tracking and an ORB-SLAM3~\cite{campos_orb-slam3_2021} implementation which aims to correct drift within the SLAM map. For this approach, the external tracking method is considered the ground truth. Our approach will never be more accurate than the external tracking method that is used. Essentially, any tracking method that corrects drift can be utilized. We are also not limited to vision-based external tracking methods. These methods are used to place the EXT anchors, which essentially consist of a 6DOF camera pose comprising both position and orientation. The pose is transmitted to the headset using a network protocol. The Open Sound Control (OSC) protocol~\cite{wright_open_1997} was employed in this case, thanks to its simplicity and support by most game engines and headsets. The following paragraphs provide further elaboration on both tracking methods. Both tracking methods employ a Leica RTC360 laser scanner to represent the real world accurately. It is important to note that this is not a requirement for our system. Other scanning methods are improving and can also be employed. 

\subsubsection{Marker-based tracking}
\begin{figure}
    \centering
    \includegraphics[width=0.8\textwidth]{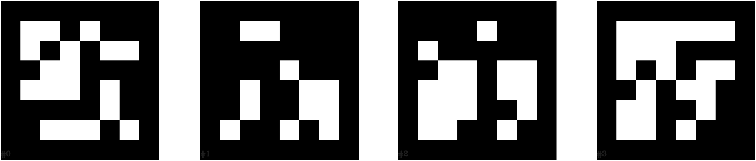}
    \caption{The ArUco markers that were used for both external tracking methods. These markers are from the aruco\_mip\_36h12 dictionary.}
    \label{fig_markers}
\end{figure}

In environments where feasible, markers can be strategically positioned for tracking purposes. The markers that were used are from the aruco\_mip\_36h dictionary, as seen in Figure \ref{fig_markers}. Each location where we wish to place an anchor is marked. The markers are printed 16 cm in size. By obtaining a laser scan of the surroundings, the precise locations of these markers can be found. Then, employing marker detection techniques outlined by Garrido-Jurado et al.~\cite{garrido-jurado_automatic_2014,garrido-jurado_generation_2016} and Romero-Ramirez et al.~\cite{romero-ramirez_speeded_2018}, alongside the known 3D positions of the markers, the camera's pose can be determined within the coordinate system of the laser scan. This approach inherently eliminates the introduction of drift, as the camera's pose is only established in relation to the markers and thus in relation to the global world. However, errors can still occur due to inaccuracies in marker detection or deviations in the laser scan. This might be due to the marker not being completely flat, sudden movements, differing lighting conditions, etc.

\subsubsection{Based on an ORB-SLAM3 implementation}
A second method is explored in situations where only a few markers are allowed, as studied by Michiels et al.~\cite{michiels_tracking_2024}. The same markers were employed as in the marker-based tracking method. This approach is based on ORB-SLAM3~\cite{campos_orb-slam3_2021} and focuses on reducing drift by using a small number of markers and a laser scan as a ground truth. In this technique, a SLAM map is constructed utilizing an external camera. Subsequently, adjustments, such as deformations and optimizations, are made to this map to align it as closely as possible with the laser scan containing markers, thereby minimizing drift. Although this method still necessitates a few markers for alignment, the inclusion of ORB-SLAM3 enables tracking even when markers are not visible. Consequently, users are not limited to placing anchors around markers. The accuracy is approximately 10 cm, which is sufficient for many virtual reality applications where the user is completely separated from the real world, such as operator training VR simulations, gaming, etc. However, AR requires higher accuracy to integrate digital content seamlessly into the physical world. The additional keyframe alignment from Michiels et al.~\cite{michiels_tracking_2024} can also be employed to improve accuracy further. 

Should a more precise tracking method emerge in the future, the system is ready to seamlessly integrate it without necessitating any adjustments. By simply incorporating the new approach, the performance of our current system can be enhanced. This means that our system's accuracy relies on the external method's precision. Another benefit is that the external tracking can operate at a slower frequency, as long as it maintains accuracy.

\subsection{Internal tracking}
Since the INT anchors are spatial anchors, they are continuously corrected for drift inside the headset. Thus, these anchors are accurate points in the tracking map, on which we can rely for registration. Spatial anchors are placed and saved inside the headset via the appropriate SDK, as shown in Table \ref{tab_sdks}.

\begin{table}
    \centering
    \begin{tabular}{c|c}
         VIVE HTC &  Wave VR SDK \\
         Meta & Oculus VR SDK \\ 
         Pico & ~~~Pico Integration SDK \\ 
         Microsoft Hololens~~~ & ~~~Mixed Reality Toolkit \\ 
         Others & ...
    \end{tabular}
    \caption{The spatial anchor SDK for different brands of headsets.}
    \label{tab_sdks}
\end{table}

Any XR headset that supports the concept of spatial anchors can be used with this approach. Spatial anchors can be shared across multiple devices, allowing for easy setup of multiple devices. 

\section{Operational stage}
Currently, our system has two sparse sets of anchors (clouds): the INT anchor cloud, which comprises spatial anchors, and the EXT anchor cloud, which encompasses camera poses within the coordinate system of the virtual world. Now, the external tracking camera is no longer necessary, as long as the headset retains the spatial anchors. We can use the XR headset in standalone mode since all the required information has already been gathered.

While the spatial anchors continuously correct for drift, ensuring the accuracy of the anchors in space, they are most effective within a range of 3 meters from their origin. To ensure alignment over a large area, it is necessary to fit the EXT anchor cloud with the INT anchor cloud. The presence of other nearby anchors diminishes the drift of anchors at larger distances. Cloud registration techniques are employed to achieve the alignment. This paper demonstrates two options for registration: weighted and non-weighted registration. The weighted version considers the user's distance to the anchor, giving more priority/weight to the anchors that are close. The non-weighted (static) version computes the best overall alignment without adjusting for the user-anchor distance. 

The goal of the registration is to find the transformation, thus rotation and translation, from one set of points to the other. Assume the point clouds $\{p_i\}$ and $\{p_i'\}$, where $i = 1,...,N$, are related as follows:
\begin{equation}
\label{cloudreg}
p_i' = R * p_i + T + N_i'
\end{equation}
where $R$ denotes a rotation matrix, $T$ a translation vector, and $N_i'$ a noise vector. The determination of the transformation can be achieved using least-squares fitting, employing singular value decomposition (SVD) as demonstrated by Arun et al.~\cite{arun_least-squares_1987}. First, both point clouds are centered around their centroid by computing the mean $\bar{p}$ of a point cloud $\{p_i\}$, where each $p_i$ is a three-dimensional vector $(x_i, y_i, z_i)$.
\begin{equation}
\label{mean}
    \bar{p} = (\frac{1}{N} \sum_{i=1}^{N} x_i, \frac{1}{N} \sum_{i=1}^{N} y_i, \frac{1}{N} \sum_{i=1}^{N} z_i)
\end{equation}
Then, the above relation (Equation \ref{cloudreg}) is minimized by calculating the 3x3 covariance matrix and decomposing it into a rotation matrix utilizing SVD. This describes the non-weighted method of cloud registration. 

The weighted point cloud registration method is a slightly modified version of the above method. Instead of computing the center of mass as the mean of the point cloud (Equation \ref{mean}), the center of mass is calculated as a function of the user-anchor distance $d_i$ as follows:
\begin{equation}
    \bar{p} = (\frac{\sum_{i=1}^{N} w_i * x_i}{\sum_{i=1}^{N} w_i},\frac{\sum_{i=1}^{N} w_i * y_i}{\sum_{i=1}^{N} w_i}, \frac{\sum_{i=1}^{N} w_i * z_i}{\sum_{i=1}^{N} w_i})~|~ w_i = 1/d_i^2
\end{equation}
A greater user-anchor distance corresponds to a lower weight, contributing less to the center of mass. Although several functions can be used to determine the weight from the user-anchor distance, we found this function to result in a smooth interpolation between the different anchors. 

Since the pose of the spatial anchors is updated on a small scale for every frame, point cloud registration is applied for every frame as well. In the case of the static registration method, this frequency could be reduced by only computing a new registration if a spatial anchor is moved further than a certain threshold. 

\section{Evaluation}
The complete system is empirically evaluated by overlaying a laser scan mesh of the environment on the passthrough layer of the headset. This shows the offset between the real world and the virtual world, giving an indication of accuracy. 

This approach aims to approximate the external tracking method as closely as possible, as it is considered the ground truth and the best achievable alignment. Thus, the resulting tracking solution inside the headset is compared with the external tracking method as a ground truth. 

The headset used for evaluation is the Meta Quest 3. Using a custom 3D-printed mount, a RealSense D455 camera sensor is connected to the headset. The resolution that is used for both tracking methods is $(640 \times 480)$. The point clouds and deduced meshes for the external tracking methods are obtained using the Leica RTC360 laser scanner. The applications were built in Unreal Engine 5.3. The system is evaluated with two applications: \textit{Utility infrastructure maintenance} and an \textit{Artistic performance}.

\subsection{App 1: Utility infrastructure maintenance}
\begin{figure}
    \centering
    \includegraphics[width=0.8\textwidth]{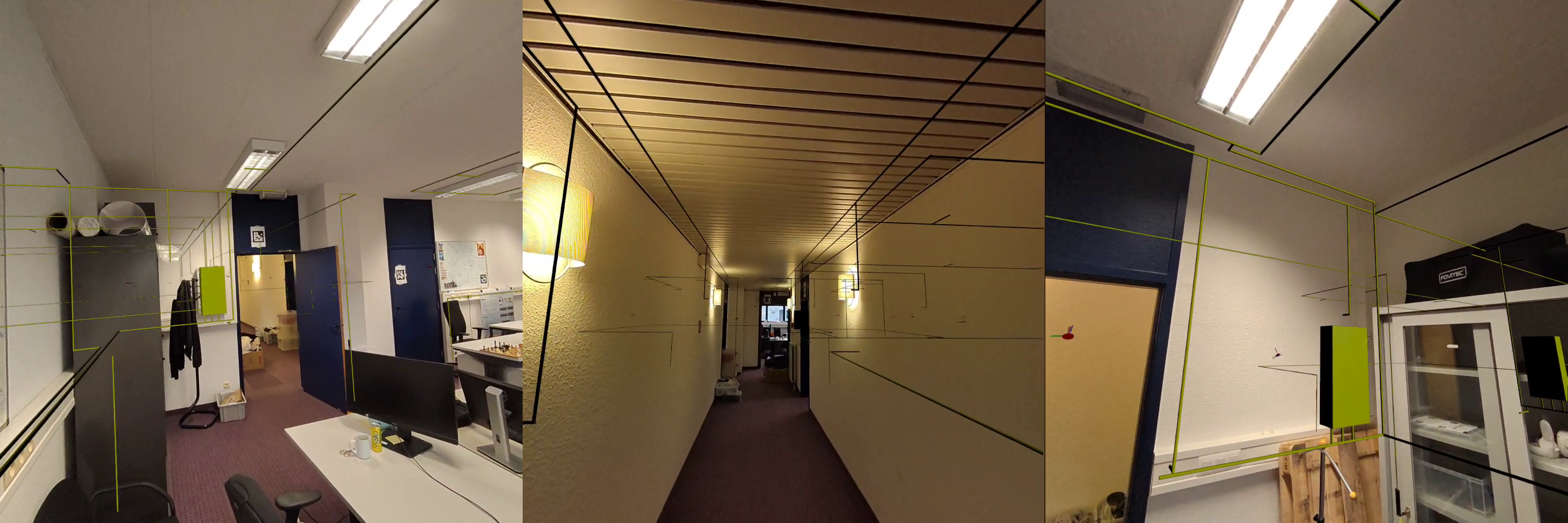}
    \caption{App 1: Utility infrastructure maintenance. Electrical circuits are displayed. The user's perspective of the Quest 3's passthrough view is displayed.}
    \label{fig_edm_img}
\end{figure}

This application shows electrical and heating infrastructure to users, which could help architects, engineers, construction workers, and maintenance personnel by visually representing the building's utility systems as seen in Figure \ref{fig_edm_img}. This type of application would enable users to accurately locate and understand the layout of utility infrastructure within a building, facilitating efficient and effective management of these systems. Having precise content registration is beneficial for these applications. Building information models (BIM) contain the necessary information for this application, including electrical circuits and Heating, Ventilation and Air Conditioning (HVAC) details. Note that a BIM model for the building in this example does not exist. Instead, the electrical wiring was modeled into an environment mesh in a speculative manner, thus inaccurate. This wiring only serves as a proof of concept. This highlights another benefit of this approach: its simplicity in content creation for a large area. This application was integrated in an environment of 16.5 m $\times$ 25 m. 

\subsection{App 2: Artistic performance}
\begin{figure}
    \centering
    \includegraphics[width=0.8\textwidth]{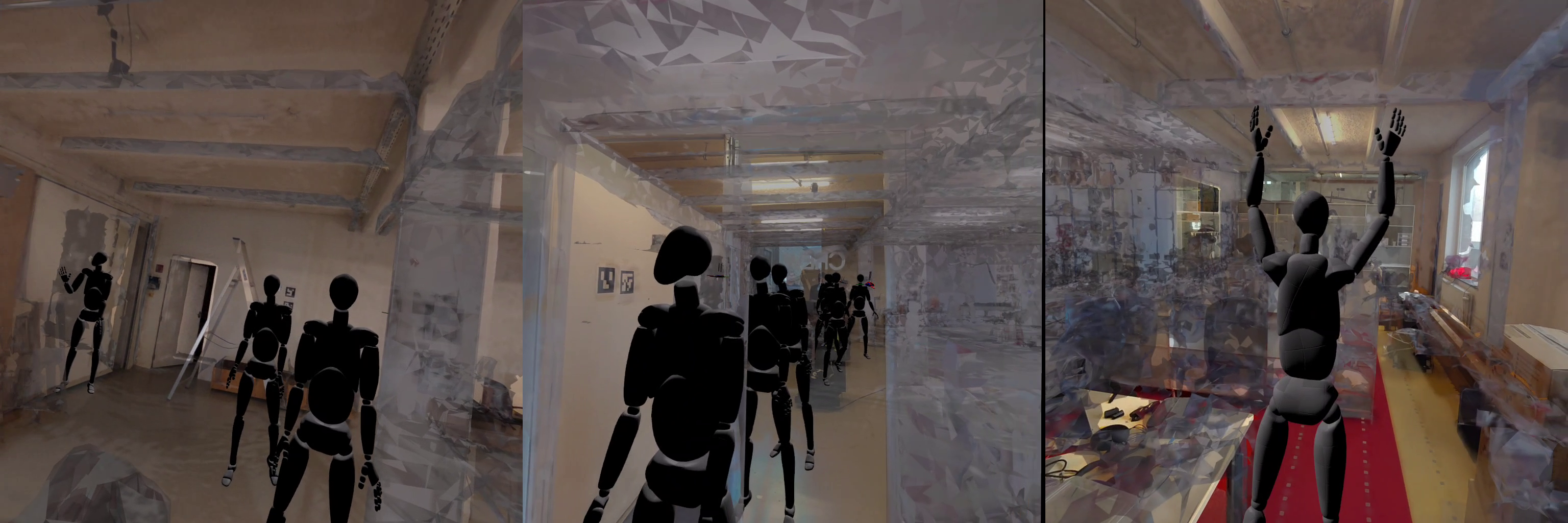}
    \caption{App 2: Artistic performance, showing the power of art in XR in a well-aligned co-located environment. A blended view of the virtual content onto the passthrough view of the Quest 3 is shown. }
    \label{fig_crew_performance}
\end{figure}

This application showcases an artistic performance in a co-located environment. Figure \ref{fig_crew_performance} shows a few images of the production. Accurate alignment with the physical world is essential for this application to be immersive and convincing. This application was tested in an environment covering 16 m $\times$ 12 m. 

\subsection{Empirical evaluation}
\begin{figure}
    \centering
    \includegraphics[width=0.8\textwidth]{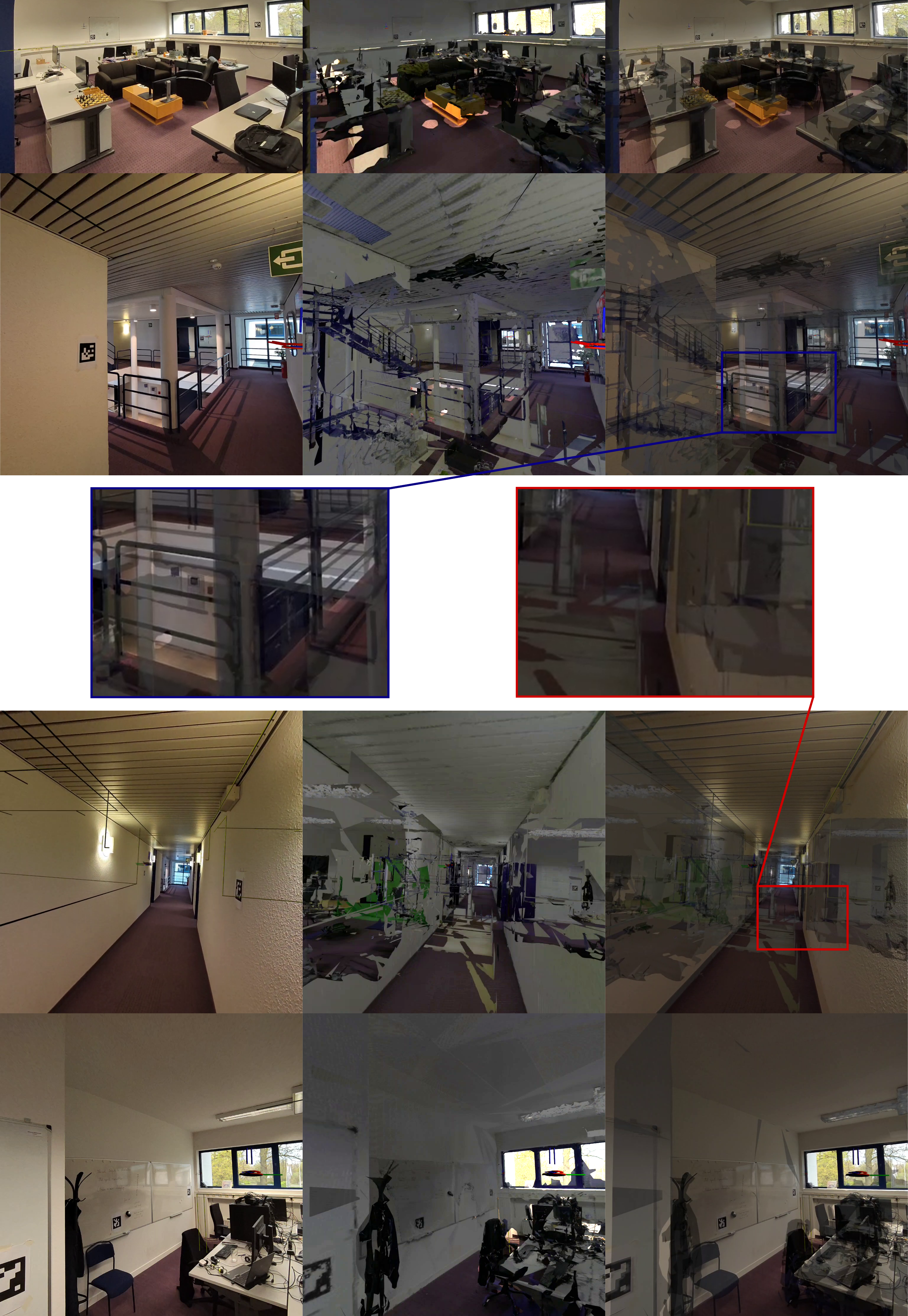}
    \caption{App 1: Virtual scene alignment with the real world using ORB-SLAM tracking in the preparation stage and a Quest 3. A passthrough view can be seen in the left column. The middle column shows the virtual environment, while the right column displays a blended view of the virtual content onto the passthrough view. There are two zoomed-in sections displayed to highlight the accuracy.}
    \label{fig_edm_orb}
\end{figure}

\begin{figure}
    \centering
    \includegraphics[width=0.8\textwidth]{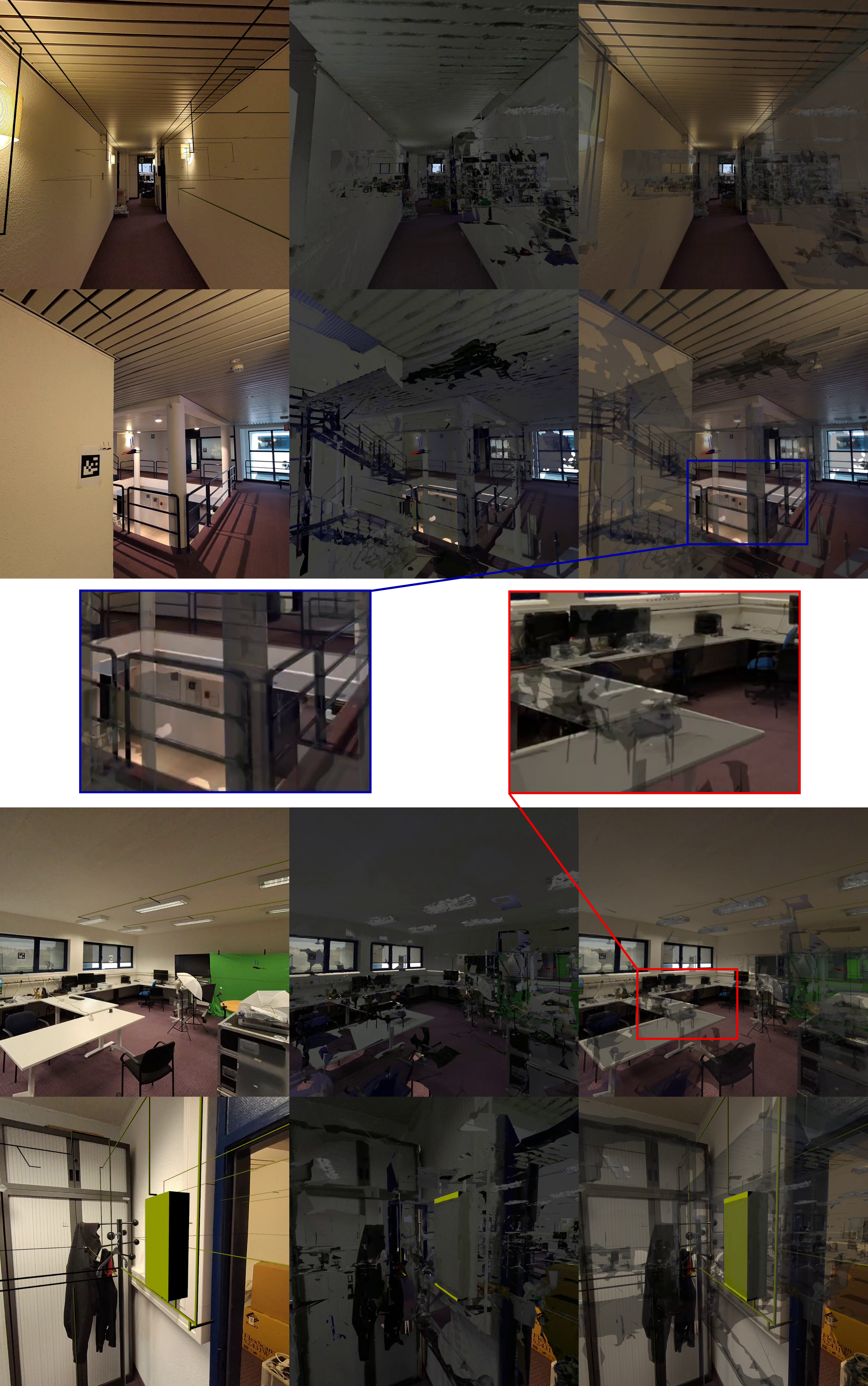}
    \caption{App 1: Virtual scene alignment with the real world using marker-based tracking in the preparation stage using a Quest 3. A passthrough view can be seen in the left column. The middle column shows the virtual environment, while the right column displays a blended view of the virtual content onto the passthrough view. There are two zoomed-in sections displayed to highlight the accuracy.}
    \label{fig_edm_emp}
\end{figure}

\begin{figure}
    \centering
    \includegraphics[width=0.8\textwidth]{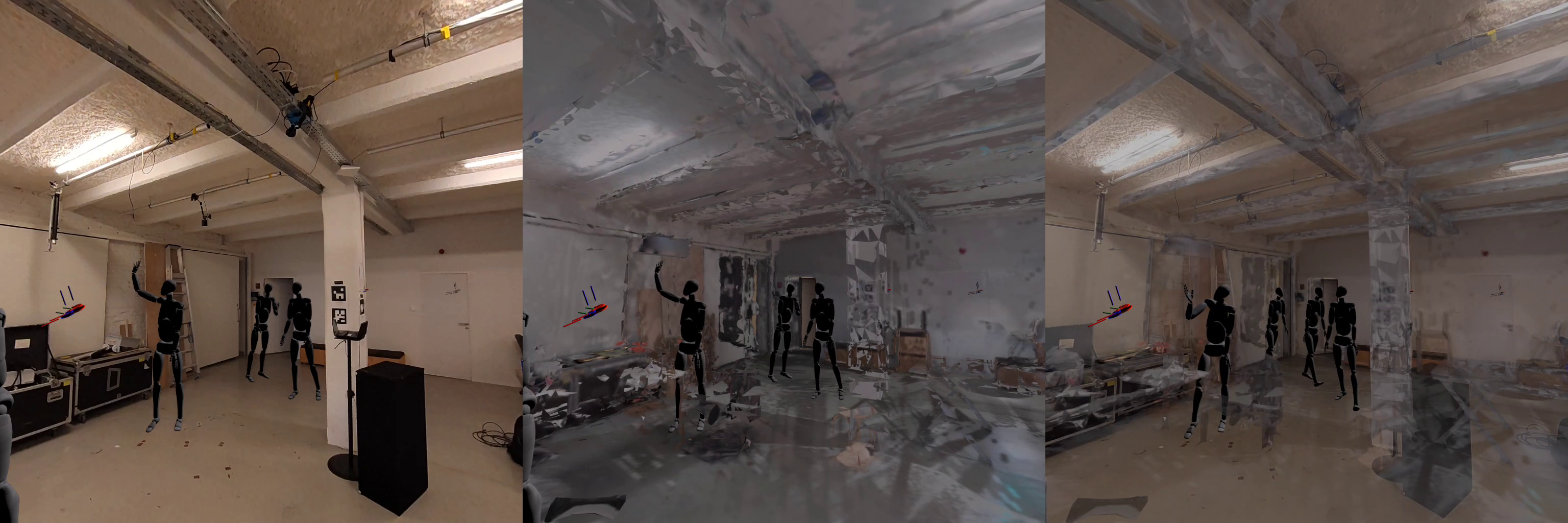}
    \caption{App 2: Virtual scene alignment with the real world using marker-based tracking in the preparation stage using a Quest 3. A passthrough view can be seen in the left image. The middle image shows the virtual environment from the same perspective, while the right image displays a blended view of the virtual content onto the passthrough view.}
    \label{fig_crew_emp}
\end{figure}

The precision of the content alignment is shown in Figures \ref{fig_edm_orb}, \ref{fig_edm_emp} and \ref{fig_crew_emp}. This empirically proves the quality of the content alignment, yet it is important to note that the mesh used, can also contain faulty parts. Figure \ref{fig_edm_emp} shows the content alignment using anchors placed with the marker-based tracking method, while Figure \ref{fig_edm_orb} shows the same environment, but with anchors placed using the ORB-SLAM3 method. As can be seen, overall the marker-based method yields better alignment compared to the ORB-SLAM3 method. This shows that a more accurate external tracking method results in a more accurate alignment.

\subsection{Quantitative evaluation}
For quantitative assessment of the system, we conducted a comparison of camera poses simultaneously obtained through both internal and external tracking methods. The marker-based tracking method was utilized to retrieve camera poses, using markers previously employed for alignment. However, these comparisons were made at least 1 meter away from the anchors. Table \ref{table_markers} presents the average error and the standard deviation of the distance and angle. These tables illustrate the error of the estimated internal camera poses in relation to the camera poses derived from the external, marker-based tracking. \textit{App 1} and \textit{App 2} display the results obtained with evenly distributed markers while \textit{App 2 with 2 markers} presents outcomes with only two markers that are approximately 11 meters apart. Each set of data includes four measurements: (\textit{static, controller}), (\textit{static, anchor}), (\textit{weighted, controller}) and (\textit{weighted, anchor}). In this context, `static' and `weighted' refer to the static and weighted cloud registration methods, while `controller' and `anchor' denote the calibration methods of the external camera to the XR headset. 

Table \ref{table_markers} does not directly display a metric for the alignment, but rather how well the internal tracking of the headset follows the external tracking method. Thus, this evaluation considers the external tracking method the ground truth. Additionally, it is important to note that inconsistencies in tracking are also included in the errors displayed below. The more consistent the pose estimation of a tracking solution, the better the results will be. We only included marker-based tracking in these results because it yields the most consistent results. 

\begin{table}
\caption{The error of the estimated internal camera poses in relation to the ground truth: the camera poses obtained from the marker-based tracking method. For \textit{App 1} and \textit{App 2}, the markers used were evenly spread. \textit{App 2 with 2 markers} shows results in which only two markers were used, approximately 11 meters apart.}
\centering
\begin{tabular}{l|cccc}
\hline
\multicolumn{1}{r|}{Metric} & \multicolumn{2}{c}{\cellcolor[HTML]{FFFFFF}Angle (°) ↓} & \multicolumn{2}{c}{\cellcolor[HTML]{FFFFFF}Distance (cm) ↓} \\
Method & \cellcolor[HTML]{FFFFFF}$\mu$ & \cellcolor[HTML]{FFFFFF}$\sigma$ & \cellcolor[HTML]{FFFFFF}$\mu$ & \cellcolor[HTML]{FFFFFF}$\sigma$ \\ \hline
App 1(static, controller) & 1.22 & 0.54 & 3.84 & 1.80 \\
App 1(static, anchor) & \textbf{1.17} & \textbf{0.48} & 3.90 & \textbf{1.72} \\
App 1(weighted, controller) & 1.23 & 0.55 & 3.27 & 2.07 \\
\textbf{App 1(weighted, anchor)} & \textbf{1.17} & 0.52 & \textbf{3.16} & 1.98 \\ \hline
\rowcolor[HTML]{FFFFFF} 
App 2(static, controller) & 0.77 & 0.30 & 3.05 & 1.56 \\
\rowcolor[HTML]{FFFFFF} 
App 2(static, anchor) & 0.97 & 0.54 & 3.22 & 1.52 \\
\rowcolor[HTML]{FFFFFF} 
\textbf{App 2(weighted, controller)} & \textbf{0.76} & \textbf{0.28} & 2.25 & 1.26 \\
\rowcolor[HTML]{FFFFFF} 
App 2(weighted, anchor) & 0.98 & 0.52 & \textbf{2.24} & \textbf{1.23} \\ \hline
 & \multicolumn{1}{l}{} & \multicolumn{1}{l}{} & \multicolumn{1}{l}{} & \multicolumn{1}{l}{} \\ \hline
App 2 with 2 markers(static, controller) & 1.23 & 0.45 & 5.5 & 2.68 \\
App 2 with 2 markers(static, anchor) & 1.35 & 0.47 & 5.71 & 2.76 \\
\textbf{App 2 with 2 markers(weighted, controller)} & \textbf{1.24} & 0.45 & \textbf{5.4} & \textbf{2.51} \\
App 2 with 2 markers(weighted, anchor) & 1.35 & \textbf{0.42} & 5.6 & 2.67 \\ \hline
\end{tabular}
\label{table_markers}
\end{table}

Table \ref{table_markers} shows that the weighted cloud registration yields the smallest errors. As can be seen, the calibration method that is used does not have a large impact on the results. In one case, anchor-based calibration outperforms the rest, while the controller-based calibration is better in the other. This indicates that both calibration methods achieve similar levels of accuracy. However, the anchor-based calibration method is more convenient and flexible, making this a more attractive option. As expected, the weighted cloud registration method proves to be more precise. This method assigns greater weight to nearby anchors, thereby improving alignment in their surroundings and effectively reducing drift. 
The results for \textit{App 2} seem to be better than the results of \textit{App 1}. For both spaces, roughly the same number of anchors were used. However, the area of \textit{App 1} is twice as large as the area of \textit{App 2}. We can only place anchors where there are markers. This also shows the impact of using less markers and thus anchors. 

Reducing the number of markers decreases accuracy as drift increases across larger distances. This is shown in Table \ref{table_markers}. For the $(weighted, controller)$, the distance error increases from 2.25 cm to 5.4 cm, while the angle increases from 0.76° to 1.24°. The standard deviation for both the angle and distance also increases. In environments where markers are impractical, alternative external tracking solutions are available. For example, the aforementioned ORB-SLAM3 adaptation could be employed, which greatly reduces the need for markers.

The accuracy of the system relies on the accurate placement of anchors and the calibration of the XR headset to the external camera. The weighted cloud registration technique is more accurate if all anchors are accurately placed. However, if a few anchors are misplaced due to errors in tracking, the static cloud registration improves alignment by reducing the impact of the misplaced anchors.

\section{Conclusion}
In conclusion, this paper has presented an approach to addressing the issue of drift within the headset for co-located XR spatial tracking systems. By leveraging spatial anchors and an external, drift-corrected tracking method, our proposed methodology offers a scalable solution to mitigate drift and improve accuracy over larger areas. 

Our approach builds upon existing research in this area, offering improvements over previous methods addressing practical challenges such as user comfort and additional equipment requirements. By enabling full standalone functionality of the headset during the operational stage, we enhance user freedom and usability. As the external tracking improves accuracy in the future, our method improves as well, without needing any adjustments. 

Future work could explore how to handle outliers and wrongly placed anchors. These outliers could be detected automatically using RANSAC \cite{fischler_random_1981}, for example. Additionally, using different types of external tracking systems could be researched. The system's compatibility with various XR devices can be expanded as well. 
\newpage

\subsubsection{Acknowledgements} This work was funded by the European Union’s
Horizon Europe Programme under grant agreement 101070072, Flanders Make’s XRTwin SBO project (R-12528), MAX-R (Mixed Augmented and eXtended Reality media pipeline), the Special Research Fund (BOF) of Hasselt University (R-14360) and a specialized FWO fellowship grant (1SHDZ24N).

%
%
\bibliographystyle{splncs04}
\bibliography{references}
\end{document}